\documentclass[longnamesfirst,apjl]{emulateapj}
\usepackage{apjfonts,natbib}
\tighten

\received{10 November 2003}
\accepted{16 January 2004}

\slugcomment{The Astrophysical Journal, Accepted}
\shortauthors{THUAN ET AL.}
\shorttitle{CHANDRA OBSERVATIONS OF SBS~0335$-$052, SBS~0335$-$052W,
AND I~ZW~18}

\begin{document}

\title{{\it Chandra} Observations of the Three Most Metal-Deficient Blue 
Compact Dwarf Galaxies known in the Local Universe, 
SBS~0335$-$052, SBS~0335$-$052W, and I~Zw~18}

\author{Trinh~X.~Thuan,\altaffilmark{1} 
Franz~E.~Bauer,\altaffilmark{2}
Polichronis~Papaderos,\altaffilmark{3} 
and
Yuri~I.~Izotov\altaffilmark{4}}

\altaffiltext{1}{Astronomy Department, University of Virginia,
P.O. Box 3818, University Station, Charlottesville, VA, 22903, USA;
txt@virginia.edu}
\altaffiltext{2}{Institute of Astronomy, University of Cambridge,
  Madingley Rd., Cambridge, CB3 0HA, UK; feb@ast.cam.ac.uk}
\altaffiltext{3}{Universit\"ats-Sternwarte, Geismarlandstrasse 11, D-37083 
G\"ottingen, Germany; papade@uni-sw.gwdg.de}
\altaffiltext{4}{Main Astronomical Observatory, National Academy of 
Sciences of Ukraine, 03680 Kyiv, Ukraine; izotov@mao.kiev.ua}

\begin{abstract}
  We present an X-ray study of the three most metal-deficient blue
  compact dwarf (BCD) galaxies known in the local Universe, based on
  deep {\it Chandra} observations of SBS~0335$-$052
  (0.025$Z_{\odot}$), SBS~0335$-$052W (0.02$Z_{\odot}$) and I~Zw~18
  (0.02$Z_{\odot}$). All three are detected, with more than 90\% of
  their X-ray emission arising from point-like sources. The 0.5--10.0
  keV luminosities of these point sources are in the range
  (1.3--8.5)$\times10^{39}$~erg~s$^{-1}$. We interpret them to be
  single or a collection of high-mass X-ray binaries, the luminosities
  of which may have been enhanced by the low metallicity of the gas.
  There are hints of faint extended diffuse X-ray emission in both
  SBS~0335$-$052 and I~Zw~18, probably associated with the
  superbubbles visible in both BCDs. The spectrum of I~Zw~18 shows a
  O~VIII hydrogen-like emission line. The best spectral fit gives an O
  overabundance of the gas in the X-ray point source 
by a factor of $\sim$ 7 with respect to the Sun, or a factor of $\sim$ 350 
with respect to the O abundance determined for the HII region.
\end{abstract}


\section{Introduction}\label{introduction}
Galaxy formation is one of the most fundamental problems in
astrophysics.  To understand how galaxies form, we need to unravel how
stars form from the primordial gas and how the first stars interact
with their surrounding environments.  As there are no heavy elements
in the early universe, the thermodynamic behavior of the gas is
essentially controlled by H$_{2}$ cooling, and the first Population
III stars are expected to be very massive \citep[e.g.,][]{Abel2002,
  Bromm2002}.  While much progress has been made in finding large
populations of galaxies at high ($z\geq3$) redshifts
\citep[e.g.,][]{Steidel1996}, truly young galaxies in the process of
forming remain elusive in the distant universe. The spectra of those
far-away galaxies generally indicate the presence of a substantial
amount of heavy elements, implying previous star formation and metal
enrichment. Instead of focussing on high-redshift galaxies, another
approach is to study massive star formation and its interaction with
the ambient interstellar medium (ISM) in a class of nearby
metal-deficient dwarf galaxies called Blue Compact Dwarf (BCD)
galaxies, some of which are thought to be undergoing their first
episode of star formation.
 
The formation of the hot gas phase is one of the most fundamental
processes operating in the early period of galaxy formation and it has
important consequences on the subsequent dynamical state and
photometric evolution of the dwarf system. The generation of large
amounts of hot (a few $\times 10^6$~K), rarefied gas is the result of
injection of energy and momentum into the cold ambient ISM by stellar
winds from massive stars and supernovae.
The starburst activity which fuels the ISM of young dwarf galaxies
with hot X-ray emitting gas lasts about 10 million years. It is then
followed
by a long ($>1$--2~Gyr) quiescent period of passive photometric 
evolution.  Moreover, expansion of the hot ISM on scales comparable to
the galactic scale length can lead to a rapid funneling of gaseous
mass into the cold gaseous halo.  The details of such a process are
expected to be sensitively related to the energy injection rate into
the ISM and the geometry and robustness of the ambient cold gaseous
medium \citep{DeYoung1994}.  Extensive mass loss can lead to an
expansion of the stellar population and to a morphological evolution
of the dwarf galaxy (for example a dwarf irregular galaxy may lose its
gas and evolve into a dwarf elliptical) or even disrupt the system
entirely \citep{Hills1980, Yoshii1987}.

Here we present a study at X-ray energies of the three most
metal-deficient star-forming dwarf galaxies known in the local
universe, the BCDs SBS~0335$-$052 \citep[][hereafter TIL]{Thuan1997b,
  Izotov1997, Thuan1997a}, SBS~0335$-$052W \citep{Lipovetsky1999}, and
I~Zw~18 \citetext{see \citealp{Papaderos2002}, \citealp{Hunt2003a},
  and references therein}. X-ray emission provides an efficient and
direct probe of the hot gas component, and can provide valuable
constraints on the nature of the X-ray binary population. The number
of high mass X-ray binaries (HMXBs) put limits on the BCD's supernova
rate while the number of low mass X-ray binaries (LMXBs) contrains the
total mass of forming stars. The relative proximity of these three BCDs
[D~$=$~54.3~Mpc for SBS~0335$-$052 and SBS~0335$-$052W (TIL);
D~$=$~12.6~Mpc for I~Zw~18 \citep{Ostlin2000}] , their extremely low
metallicities [0.025Z$_\odot$ for SBS~0335$-$052 \citep{Izotov1999a},
0.020Z$_\odot$ for SBS~0335$-$052W \citep{Lipovetsky1999}, and
0.020Z$_\odot$ for I~Zw~18 \citep{Izotov1999a}], and high neutral gas
content \citetext{see \citealp{Pustilnik2001} for SBS~0335$-$052 and
  SBS~0335$-$052W and \citealp{vanZee1998} for I~Zw~18} make them the
best local approximations to primordial young galaxies. Their
relatively bright apparent magnitudes ($m_B=17.0$~mag for
SBS~0335$-$052, $m_B=19.4$~mag for SBS~0335$-$052W and $m_B=15.8$~mag
for I~Zw~18) make them much easier to study than the very faint and
small building-block dwarf galaxies at high-redshift.  HI VLA mapping
\citep{Pustilnik2001} shows SBS~0335$-$052 and SBS~0335$-$052W to be
embedded in a common very large HI envelope with an overall size of
$\sim$~66$\times$22~kpc. There are two prominent HI peaks separated in
the east-west direction by 22 kpc (84\arcsec). The eastern peak is
associated with SBS~0335$-$052, while the western peak is about a
factor of 1.3 brighter in the HI line and is associated with the
fainter SBS~0335$-$052W.  Likewise, HI VLA mapping of I~Zw~18 by
\citet{vanZee1998} shows that it is also embedded in an extended HI
envelope.

Optical {\it HST} imaging shows that nearly all star formation in
SBS~0335$-$052 occurs in six very blue super-star clusters (SSCs) with
absolute $V$ luminosities between -11.7 and -14.7 mag, and surface
brightnesses $\ga$~100 times those of clusters and associations in
normal HII regions. The SSCs are confined to a region of about 520~pc
in diameter (TIL).  There is a systematic color gradient from the
brightest and bluest SSC at one end to the faintest and reddest SSC at
the other end. Such behavior can be attributed partly to variable dust
extinction.  Dust is clearly present, even in such a metal-poor
environment, as evidenced by the strong mid-infrared emission of
SBS~0335$-$052 \citep{Thuan1999}.  However, most of the
color gradient is due to a systematic aging of the SSCs resulting from
sequential propagating star formation (with a velocity $\leq$ 20 km
s$^{-1}$) from the reddest and oldest SSC to the bluest and youngest
one \citep{Papaderos1998} .  From the $V-I$ colors and the models of
\citet{Leitherer1995}, the age of the youngest SSC is $\sim$ 4 Myr
while that of the oldest SSC is $\sim$ 30 Myr.  On larger scales, the
light of SBS~0335$-$052 is dominated by a patchy and filamentary very
blue low surface brightness (LSB) emission.  Spectroscopic studies
\citep{Izotov1997} have revealed that ionized gas contributes $\sim
30$\% of the emission from the LSB component, out to scales of $\sim
3$ kpc. The remaining $\sim 70$\% of the light comes from an
underlying stellar population which \citet{Papaderos1998} 
have shown
is not older than $\sim 500$ Myr.  A striking signature of the
collective action of massive stars and SNe on the ISM of SBS\ 
0335$-$052 is a large supershell with radius $\sim$ 380 pc to the
North of the SSCs. TIL estimate the star formation rate in
SBS~0335$-$052 from its H$\alpha$ luminosity to be $\sim$ 0.4
M$_\odot$ yr$^{-1}$.

SBS~0335$-$052W has been studied in detail by \citet{Lipovetsky1999}.
It has the same redshift as SBS~0335$-$052 and consists of at least  
three stellar clusters. It has also a very low metallicity of $\sim$ 2\%
solar and is about 2.4 magnitudes fainter than SBS~0335$-$052. 
Its very blue $R-I$ colors also suggest a very young age 
of less than $\sim 500$ Myr. 

I~Zw~18 is composed of two bright knots of star formation separated by 
5$\arcsec$ and referred to as the brighter northwest (NW) and fainter 
southeast (SE) components \citep{Papaderos2002}. Star formation proceeds 
differently in I~Zw~18 as compared to SBS~0335$-$052. There are no SSCs in 
I~Zw~18 and the derived SFR is one order of magnitude smaller ($\sim$ 
0.04 M$_\odot$ yr$^{-1}$) than in SBS~0335$-$052. \citet{Papaderos2002} and
\citet{Hunt2003a}, using deep optical and 
near-infrared imaging, have put an upper limit of 
$\sim$ 500 Myr for the age of I Zw 18.    

In $\S$\ref{reduction} we describe the properties of the X-ray sources
and their optical properties.  In $\S$\ref{discussion} we summarize
our findings and discuss the X-ray properties of SBS~0335$-$052 and
I~Zw~18 in the broader context of star formation in a low-metallicity
environment.

\begin{figure*}
\vspace{-0.0in}
\centerline{
\includegraphics[height=5.5cm]{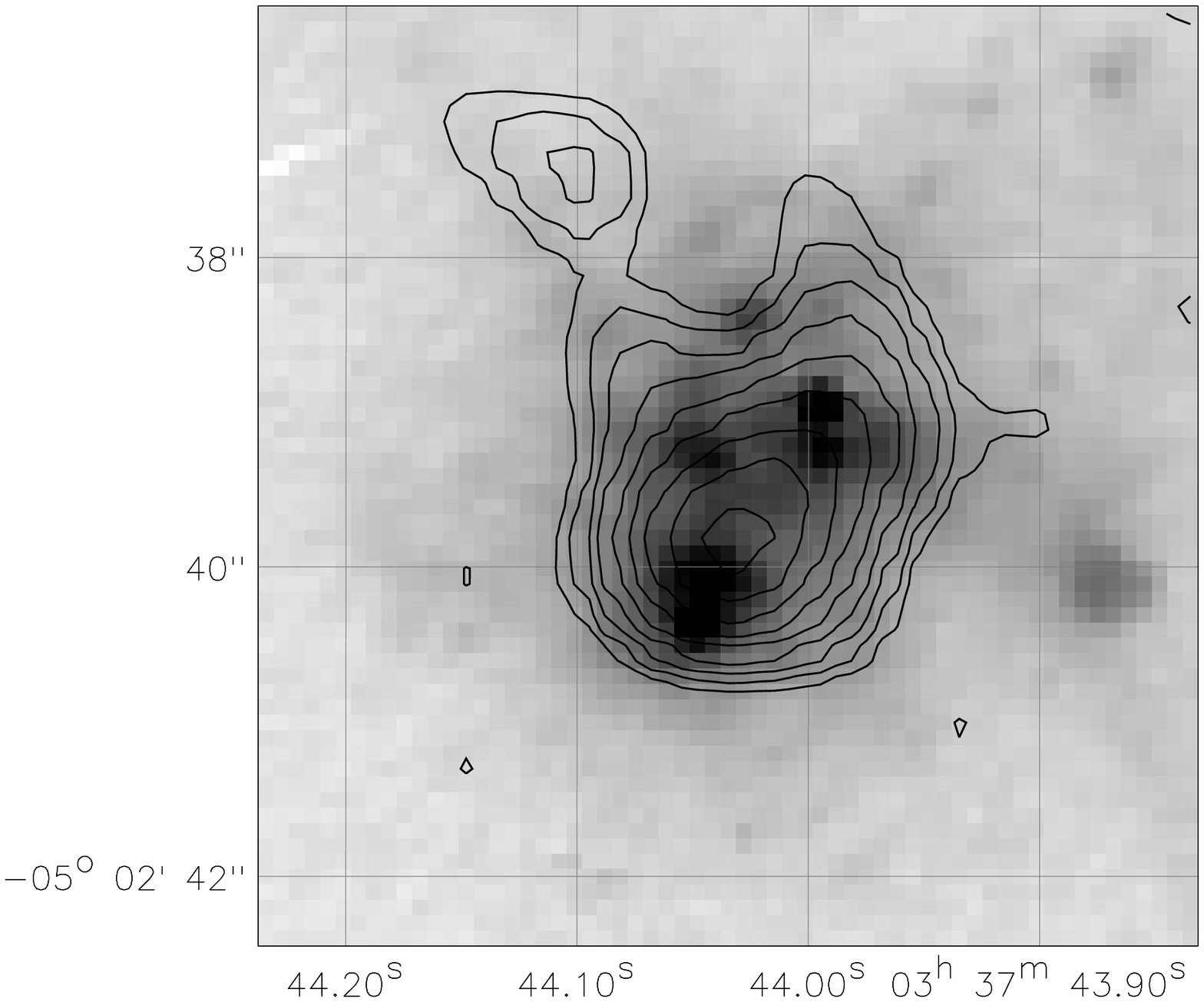}
\hglue-0.5in{\includegraphics[height=5.5cm]{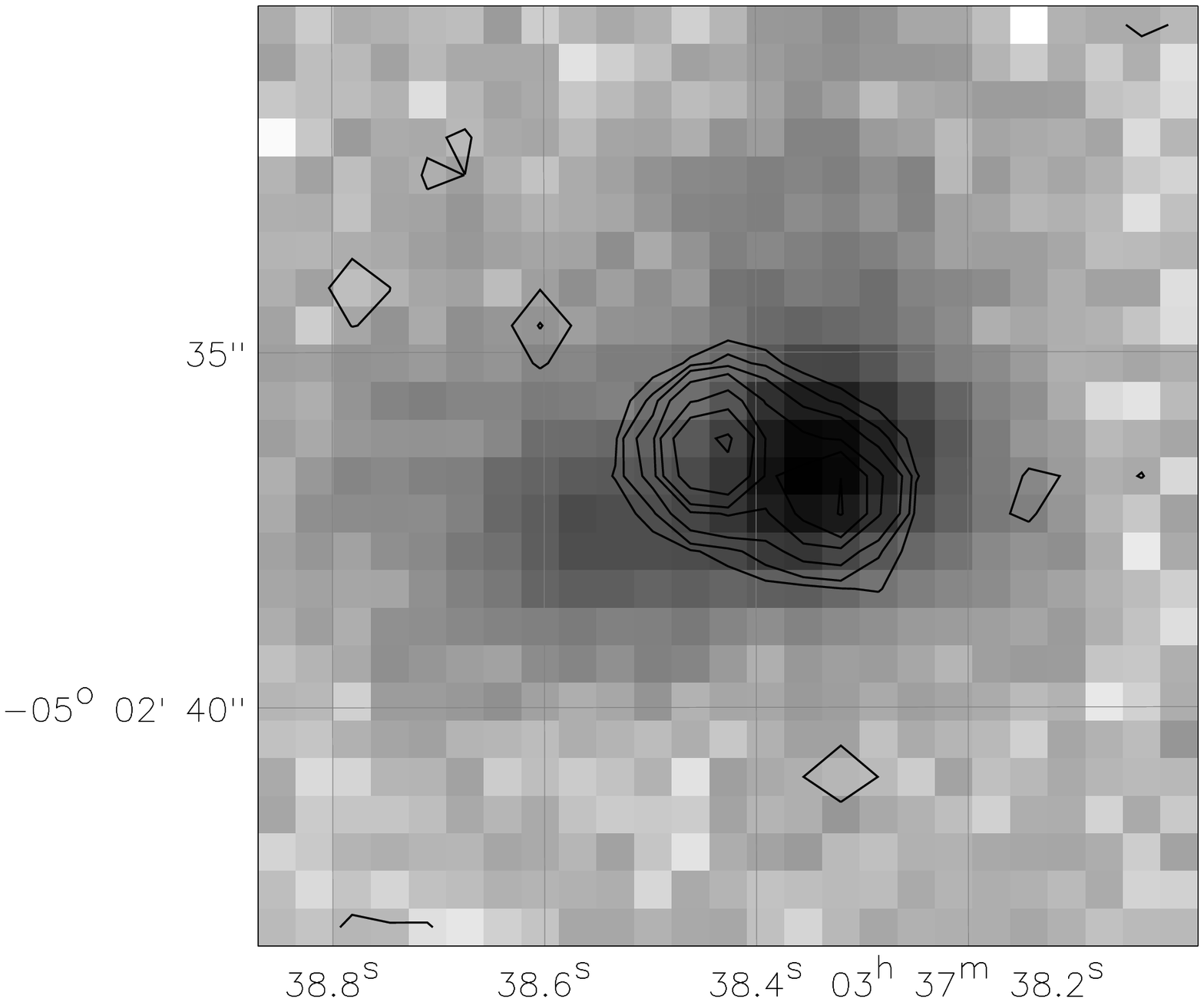}}
\hglue-0.5in{\includegraphics[height=5.5cm]{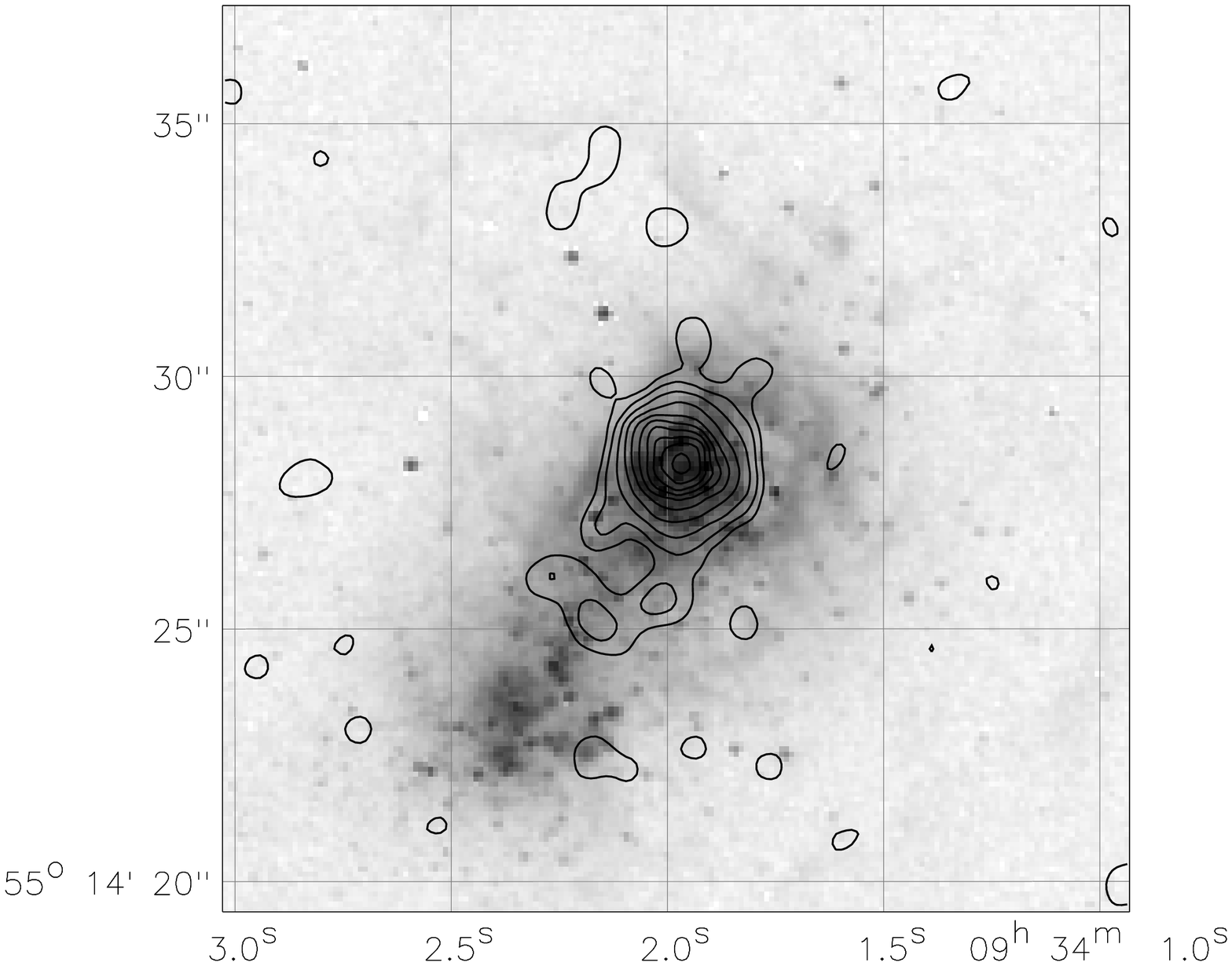}}}
\vspace{-0.2cm} 
\figcaption{X-ray contours overlaid on a {\it HST} WFPC2 F569W image
  (TIL) of SBS~0335$-$052 ({\it left}), a $B$-band Calar Alto
image \citep{Papaderos1998} of SBS~0335$-$052W({\it middle}) and a HST
WFPC2 F555W image \citep{Papaderos2002} of I~Zw~18 ({\it
right}). \label{fig:overlay}}
\vspace{-0.3cm}
\end{figure*} 

\section{Observations and Data Reduction}\label{reduction}

\subsection{X-ray Observations}\label{xobs}

SBS~0335$-$052 was observed on 2000 September 7 with ACIS-I \citep[the
imaging array of the Advanced CCD Imaging Spectrometer;][]{Garmire2002}
aboard the {\it Chandra}  satellite.  The field of view of 16\farcm9
$\times$ 16\farcm9 was large enough to include also SBS~0335$-$052W.
Events were telemetered in Very Faint mode, and the CCD temperature
was $-120^{\circ}$~C. To compare the X-ray properties of
SBS~0335$-$052 with those of another extremely metal-deficient BCD, we
have retrieved from the {\it Chandra} archive the data for the BCD
I~Zw~18.  This galaxy was observed on 2000 February 8 (P.I: D.J.
Bomans) with ACIS-S (chips 235678 on, 8\farcm5 $\times$ 8\farcm5 per
chip). Events were telemetered in Faint mode, and the CCD temperature
was $-120^{\circ}$~C. The ACIS pixel size is 0\farcs492 while the
half-energy radius of an on-axis point source is
$\approx$~0\farcs42.

Analysis was done using primarily the {\tt CIAO} V2.3 software
provided by the {\it Chandra} X-ray Center (CXC), but also with {\tt
FTOOLS} and custom software. We removed the $0\farcs5$ pixel
randomization, cleaned the ACIS background using Very Faint mode
screening for SBS~0335$-$052, performed standard {\it ASCA} grade
selection, and excluded bad pixels and columns. The total net exposure
times were 61,042~s for SBS~0335$-$052 and 40,776~s for I~Zw~18.  The
average quiescent background rates for SBS~0335$-$052 and I~Zw~18 were
found to be 0.3 and 0.96 counts s$^{-1}$ per chip, respectively, in
good agreement with ACIS-I and ACIS-S calibration measurements. The
observation of SBS~0335$-$052 contained four background ``flares'',
the intensity of each being about three times the quiescent level and
lasting for 5--10\% of the total exposure. The observation of I~Zw~18
contained one flare about ten times the quiescent level lasting for
20\% of the total exposure \footnote{See
  http://cxc.harvard.edu/contrib/maxim/bg/index.html}. None of the
flares were strong enough to warrant removal. The resulting average
background level is thus $\approx$~0.02 counts pixel$^{-1}$ and
$\approx$~0.06 counts pixel$^{-1}$, respectively.

\subsection{Astrometry of optical and X-ray images}\label{optobs}

To compare the X-ray emission with the optical emission, we have
overlayed the X-ray contours over the $V$-band (F569W) {\it HST} WFPC2
image of SBS~0335$-$052 obtained by TIL and the $V$-band (F555W) {\it
  HST} WFPC2 image of I~Zw~18 retrieved from the archives and
reprocessed by \citet{Papaderos2002}. As SBS~0335$-$052W does not lie
within the field of view of any {\it HST} image, we have used for
optical comparison the CCD $B$-band image obtained by
\citet{Papaderos1998} with the 2.2 m Calar Alto telescope. The Calar
Alto image has a large field of view ($\sim$ 17$\arcmin$ $\times$
17$\arcmin$), and contains enough stars for aligning the {\it HST} and
X-ray images to the same astrometric reference frame. A similar
alignment was performed for I~Zw~18 using the CCD $B$-band image
obtained by \citet{Papaderos2002}, also with the 2.2 m Calar Alto
telescope. To align the optical images, sources were first extracted
from the Calar Alto images using Sextractor
\citep[v2.2.1;][]{Bertin1996}. Optical sources from the Calar Alto
images were matched to the GSC2.2 catalog for absolute astrometric
alignment, giving a coincidence of 28 sources within a 3$\sigma$
radius of 0\farcs57 and a rms scatter of 0\farcs25 for SBS~0335$-$052,
and 25 sources within a 3$\sigma$ radius of 0\farcs62 and a rms
scatter of 0\farcs23 for I~Zw~18. In the same manner, optical sources
from the Calar Alto image were used to align the {\it HST} images
\citep[TIL;][]{Papaderos2002} to the same reference frame; 51 sources
were matched within a 0\farcs82 radius and a rms scatter of 0\farcs31
for SBS~0335$-$052, while 21 sources were matched within a 0\farcs49
radius and a rms scatter of 0\farcs19 for I~Zw~18.

To align the Chandra image, we first extracted the X-ray sources using
the {\tt CIAO} wavelet algorithm {\sc wavdetect} on the 0.5-8.0~keV
images with a probability threshold of $10^{-7}$ \citep{Freeman2002},
which is equivalent to the detection of approximately one false source
over the entire ACIS array. Eighty sources were detected in
total in the SBS~0335$-$052 field. 
Positions and counts for the sources were determined using
fixed circular apertures with 95\% encircled energy radii (based on
point spread functions (PSF) derived from {\sc mkpsf} and the PSF
library). Sixty-four sources lie within the region imaged by
\citet{Papaderos1998} around SBS~0335$-$052, of which 18 X-ray sources
have plausible optical counterparts within a 3$\sigma$ radius of
0\farcs9, including both SBS~0335$-$052 and SBS~0335$-$052W.
Less than 2 false matches are expected.  Aligning the X-ray image to
the optical astrometric reference frame based on these SBS~0335$-$052
X-ray/optical matches, we find that the astrometric accuracy of the
X-ray positions has a rms scatter of 0\farcs42. Likewise, 16 sources
lie within the region imaged by \citet{Papaderos2002} around I~Zw~18,
but only four X-ray sources have plausible optical counterparts
within a 3$\sigma$ radius of 1\farcs7, including I~Zw~18. This paucity
of matches is likely due to the relatively shallow depth of the
$B$-band Calar Alto image. 
Less than 1 false match is expected.  Aligning the X-ray image to
the optical astrometric reference frame based on these I~Zw~18
X-ray/optical matches, we find that the astrometric accuracy of the
X-ray positions has a rms scatter of 0\farcs55.

\subsection{X-ray Analysis}\label{xanal}

Based on our astrometric solutions, we present X-ray contours of
SBS~0335$-$052, SBS~0335$-$052W, and I~Zw~18 in Figure~\ref{fig:overlay},
overlaid on optical images.  SBS~0335$-$052
consists of a faint X-ray point source (29 counts) lying $\approx$0\farcs3
north of SSC 2.  We follow here the SSC numbering scheme of
TIL. Because the X-ray position is no more accurate
than 0\farcs42, the data is consistent with the X-ray
source being physically associated with SSC 2. The elongated shapes of
the faint X-ray contours to the north suggest that there may also be
X-ray emission associated with SSC 3 and the pair of SSC 4 and
5. There is also a very faint X-ray source to the northeast
which appears to be associated with the supernova cavity seen in the 
optical image (TIL).  Excluding the source near SSC 2, the
total X-ray emission of the fainter features is $\la$23 counts (Table
1), making it virtually impossible to distinguish between point-like
and truly extended emission.
    
\begin{deluxetable*}{llrrrrrrrl}
\tabletypesize{\scriptsize}
\tablewidth{0pt}
\tablecaption{X-ray Emission from SBS~0335$-$052, SBS~0335$-$052W, and I~Zw~18\label{tab:sources}} 
\tablehead{
\colhead{(1)} & 
\colhead{(2)} & 
\colhead{(3)} & 
\colhead{(4)} & 
\colhead{(5)} & 
\colhead{(6)} & 
\colhead{(7)} & 
\colhead{(8)} & 
\colhead{(9)} & 
\colhead{(10)} \\
\colhead{Source} & 
\colhead{Position} & 
\colhead{Counts} & 
\colhead{Model} & 
\colhead{$N_{\rm H}$} & 
\colhead{$\Gamma$/$kT$} & 
\colhead{Fit/DOF}& 
\colhead{$F_{\rm X}$} & 
\colhead{$L_{\rm X}$} & 
\colhead{Comments} \\
}
\startdata
SBS~0335$-$052  & 033744.1-050239.5   &   29.3 $\pm$  6.5 & POW & 6.8 ($<16.3$)          & 2.1$^{+1.5}_{-1.2}$ & 24.8/24 &  6.1 & 3.5 & Point Src. \\
                &                     &                   & RAY & 5.9$^{+6.3}_{-5.4}$    & 3.6 ($>1.2$)        & 24.7/24 &  5.2 & 2.8 & \\
                &                     &                   & POW & 7.0 (fixed)            & 2.2$^{+0.6}_{-0.8}$ & 24.8/25 &  5.7 & 3.5 & \\
                &                     &                   & RAY & 7.0 (fixed)            & 2.7$^{+16.6}_{-1.3}$& 24.6/25 &  4.5 & 2.8 & \\
                & 033744.1-050239.5B  &    8.4 $\pm$  5.0 & RAY & 7.0 (fixed)            & 1.0 (fixed)         &         &  0.6 & 0.64 & Extended \\
\hline 
SBS~0335$-$052W & 033738.5-050236.5   &   82.4 $\pm$ 10.2 & POW & 5.2$^{+3.3}_{-2.7}$    & 2.8$^{+0.9}_{-0.8}$ & 41.1/56 & 10.3 & 8.5 & Point Src. \#1\\
                &                     &                   & RAY & 3.1$^{+2.3}_{-1.9}$    & 2.0$^{+2.2}_{-0.8}$ & 41.6/56 &  9.6 & 5.2 & \\
                & 033738.4-050237.3   &   36.4 $\pm$  7.1 & POW & 2.3 ($<7.1$)           & 1.9$^{+1.1}_{-0.8}$ & 21.9/30 &  6.3 & 2.8 & Point Src. \#2\\
                &                     &                   & RAY & 1.3 ($<3.0$)           & 5.4 ($>1.9$)        & 22.0/30 &  5.9 & 2.4 & \\
\hline
I~Zw~18           & 093401.9+551428.4A  &  469.5 $\pm$ 21.7 & POW & 1.44$^{+0.38}_{-0.37}$ & 2.01$^{+0.14}_{-0.16}$ & 18.1/20\tablenotemark{*} & 72.1 & 1.6 & Point Src., 0.65~keV line? \\
                &                     &                   & RAY & 0.87$^{+0.27}_{-0.24}$ & 4.06$^{+1.84}_{-1.19}$ & 23.0/20\tablenotemark{*} & 66.6 & 1.4 & \\
                &                     &                   & VRAY & 0.94$^{+0.35}_{-0.24}$ & 4.28$^{+2.25}_{-1.31}$ & 8.1/19\tablenotemark{*} & 70.4 & 1.5 & $Z^{\rm O}=7.0^{+12.2}_{-4.3}Z^{\rm O}_{\odot}$\\
                & 093401.9+551428.4B  &   22.9 $\pm$  6.9 & RAY & 1.31 (fixed)           & 1.0 (fixed)            &         &  2.0 & 0.053 & Extended\\
\enddata
\tablecomments{
Column 1: Source name.
Column 2: Source position given as CXOU~JHHMMSS.S+DDMMSS.S. 
Column 3: Background-subtracted 0.5--8.0~keV counts accumulated over
60.1~ks (SBS~0335$-$052) and 40.8~ks (I~Zw~18). Aperture photometry was
performed using 95\% encircled-energy radii for 1.5~keV for point
sources, and individual background regions were selected adjacent to
each source as noted in $\S$\ref{reduction}. The standard deviation
for the source and background counts are computed following the method
of \citet{Gehrels1986} and are then combined following the ``numerical
method'' described in $\S$1.7.3 of \citet{Lyons1991}.
Column 4: Spectral model used to fit data. POW indicates an absorbed
power-law model, whereas RAY (VRAY) indicates an absorbed Raymond-Smith
thermal plasma model (with variable O abundance)
\citep{Raymond1977}.
Columns 5 and 6: Neutral hydrogen absorption column density ($N_{\rm
H}$) in units of $10^{21}$ cm$^{-2}$. Photon index ($\Gamma$) or
thermal plasma temperature ($kT$ in units of keV) as determined from
the best-fit absorbed power-law or thermal plasma models to the ACIS
spectra. Also listed are the 90\% confidence errors calculated for one
parameter of interest ($\Delta\chi^2 = 2.7$).
Column 7: Goodness of fit/degree of freedom. For SBS~0335$-$052,
fitting was performed with the c-statistic, while for I~Zw~18 the
$\chi^2$ statistic was used (denoted by ``*'').
Columns 8 and 9: Observed 0.5--10.0~keV fluxes in
units of 10$^{-15}$~erg~cm$^{-2}$~s$^{-1}$ and absorption-corrected
0.5--10.0~keV luminosities in units of 10$^{39}$~erg~s$^{-1}$, assuming
the best-fit model parameters given in Columns 5 and 6.
Column 10: Comments.}
\vspace{-0.2cm}
\end{deluxetable*}

Although SBS~0335$-$052W is 2.4 mag fainter in $B$ than SBS~0335$-$052,
it is 2.8 times brighter in the X-ray. It consists of two faint X-ray point
sources. The fainter one appears to be spatially coincident with the
brightest of the two optical clusters in SBS~0335$-$052W. The
brighter X-ray source lies $\approx$~1\farcs8 to the northeast and 
does not appear to have an optical counterpart. 

I~Zw~18 is only 1.2 mag brighter in $B$ than SBS~0335$-$052, 
but its X-ray flux is more than 10 times higher than that of SBS~0335$-$052.
 It consists of a single X-ray point
source (471 counts) and a very faint extended X-ray component (22
counts). The point source is spatially coincident with the bright
NW component of I~Zw~18. The faint extended component appears
to lie along the main body of I~Zw~18 with some X-ray emission centered
on the northwest component and some in between the northwest and
southeast optical peaks. \citet{Bomans2002} in a preliminary
reduction of the same Chandra data also found the X-ray emission of
I~Zw~18 to be composed of a strong point source and fainter extended
emission, although we fail to recover the SW and NE extensions these
authors claim to see.

X-ray variability is an extremely useful method for constraining the
size of the emitting region. Because {\it Chandra} records the
position, energy, and time of all incoming photons, we can use this
information to search for short-term variability over the duration of
our observation.  We searched for variability from the different X-ray
components of SBS~0335$-$052, SBS~0335$-$052W, and I~Zw~18 using the
Kolomogrov-Smirnov (KS) test. While no significant variability was
found, the limited statistics do not strongly constrain the temporal
nature of the sources.

X-ray spectra can provide information about the mechanisms by which
the X-ray emission is generated.
The X-ray spectra were analyzed using {\tt XSPEC}
\citep[v11.2][]{Arnaud1996}. Unless stated otherwise,
spectral parameter errors are for the 90\% confidence level, assuming
one parameter of interest. The X-ray fluxes and absorption-corrected
luminosities for all sources were calculated from spectral fitting
using {\tt XSPEC}. Importantly, none of the sources of interest suffer
from pileup. 

To constrain the fit of the X-ray spectra, it is important to use
accurate HI column densities for each of the objects.  The HI column
density towards SBS~0335$-$052 has been derived by \citep{Thuan1997b}
through fitting the wings of the damped L$\alpha$ profile in a {\it
  HST/GHRS} UV spectrum. They found $N_{\rm HI} =
(7.0\pm0.5)\times10^{21}$~cm$^{-2}$.  This is nearly 10 times larger
than the peak HI column density of 7.4$\times10^{20}$~cm$^{-2}$ which
\citep{Pustilnik2001} found in their {\it VLA} HI map of SBS
0335--052.  This large difference can be undertood as a beam-smearing
effect in the VLA map. The synthesized VLA beam
(20\farcs5~$\times$~15\farcs0) is considerably larger than the GHRS
aperture (2\farcs0~$\times$~2\farcs0), and column densities in the
direction of structures smaller than the VLA beam will be artificially
diminished. Thus, for the X-ray point source associated with SBS
0335--052, the HI column density obtained from the spectrum through
the small GHRS aperture is the appropriate one to use. No comparable
HST/GHRS spectrum exists for SBS~0335$-$052W. The VLA map
\citep{Pustilnik2001} shows that the HI column density towards
SBS~0335$-$052W to be 1.35 larger as compared to SBS~0335$-$052. For
lack of better information, we have simply scaled the GHRS HI column
density towards SBS~0335$-$052 by that factor to obtain the HI column
density towards SBS~0335$-$052W. However, we should remember that,
because of the large VLA beam, we do not know the exact value of N(HI)
in the inner 2$\arcsec$ of SBS~0335$-$052W, where the X-ray emission
is located.  For the HI column density towards I~Zw~18, we have
adopted the value $N_{\rm HI} = (3.5\pm0.5)\times10^{21}$~cm$^{-2}$
derived by \citep{Kunth1994} in the same way as for SBS 0335--052, by
fitting the wings of the damped L$\alpha$ profile in a {\it HST/GHRS}
spectrum of the NW component, obtained through a
2\farcs0~$\times$~2\farcs0 aperture. This value is entirely consistent
with the value of 2.1$\times10^{21}$~cm$^{-2}$ derived by
\citet{VidalMadjar2000} from fitting the Ly$\beta$ absorption profile,
with that of $(2.0\pm0.5)\times10^{21}$~cm$^{-2}$ obtained by
Lecavelier des Etangs et al. (2003) and that of
$2.2^{+0.6}_{-0.5}\times10^{21}$~cm$^{-2}$ obtained by Aloisi et al.
(2003) by fitting the blue wing of the Ly$\beta$ line and the profiles
of several other lines of the HI Lyman series. The last three
determinations are based on {\it FUSE} spectra obtained through a
30\farcs0~$\times$~30\farcs0 aperture.  \citet{Brown2002} have
obtained {\it STIS} data of I Zw 18 at the very high spatial
resolution of 0\farcs5. They discovered significant inhomogeneity in
the HI gas, with a peak as high as $N_{\rm HI} \sim
2\times10^{22}$~cm$^{-2}$. However, this is not the appropriate value
to use here as this HI peak is associated with the SE component, while
the X-ray source is associated with the NW component which
\citep{Kunth1994} observed.  VLA HI maps of I Zw 18 have been obtained
by van Zee et al. (1998).  They also found a peak HI column density at
the SE component of $\sim3\times10^{21}$~cm$^{-2}$. Again, because the
highest resolution map of van Zee et al. (1998) has a beam size of
5\farcs2~$\times$~4\farcs8, beam smearing effects have decreased the
real HI peak column density by a factor of nearly 10.

The ACIS spectra of the sources in SBS~0335$-$052, SBS~0335$-$052W,
and I~Zw~18 were fit with two types of models, absorbed thermal plasma
models \citep[{\it zwabs$+$raymond};][]{Raymond1977} and absorbed
power-law models ({\it zwabs$+$pow}). Given the limited counting
statistics, we performed the spectral analysis using the unbinned,
background-subtracted source spectra and the Cash statistic
\citep{Cash1979}, which is well suited to low-count sources
\citep[e.g.,][]{Nousek1989}. Note that we expect $\la$~2 background
counts in the source aperture, so spurious residuals from spatially
varying background should be negligible. One limitation of the Cash
statistic is that it does not provide a reliable quality-of-fit
criterion (like the $\chi^2$ statistic) to compare different models.
Therefore, we established the quality of the spectral results via
visual inspection using the binned spectrum.  None of the sources have
enough counts to constrain strongly both the column density and the
temperature or photon index.
In our spectral fits using a thermal plasma model, we have assumed an
abundance of $Z=0.025 Z_{\odot}$ for SBS~0335$-$052, and $Z=0.02 Z_{\odot}$ 
for SBS~0335$-$052W and I~Zw~18, in accordance with the abundances
determined spectroscopically \citetext{see
\citealp{Izotov1999a} for SBS~0335$-$052 and I~Zw~18, and
\citealp{Lipovetsky1999} for SBS~0335$-$052W}.

The spectrum of the point source in SBS~0335$-$052 has only
$\approx$29 counts and is not strongly constrained by either the
absorbed power-law or thermal plasma models when the column density
$N_{\rm H}$ is left as a free parameter in the fit (see
Table~\ref{tab:sources}). If we fix $N_{\rm H}$ to the HI value of
$7\times10^{21}$~cm$^{2}$ obtained from fitting the Ly$\alpha$ profile
\citep{Thuan1997b}, we find that the spectrum is moderately soft, with
best-fit values of the photon index $\Gamma=2.2^{+0.6}_{-0.8}$
($\Gamma$ = $\alpha$ + 1, where $\alpha$ is the slope of the power-law
fit ) or $kT=2.7^{+16.6}_{-1.3}$ keV. These values are consistent with
those of black-hole X-ray binaries \citep[e.g.,
$\Gamma\sim1.7$--2.2;][]{Foschini2002, Roberts2003}, but are not
consistent with a hot gas component heated by supernovae which has
typically kT $<1$ keV. The point source has an unabsorbed X-ray
luminosity of (2.8--3.5)$\times10^{39}$~erg~s$^{-1}$.
The spectrum and lack of variability do not provide further constraints on the
nature of this source.

\begin{figure*}
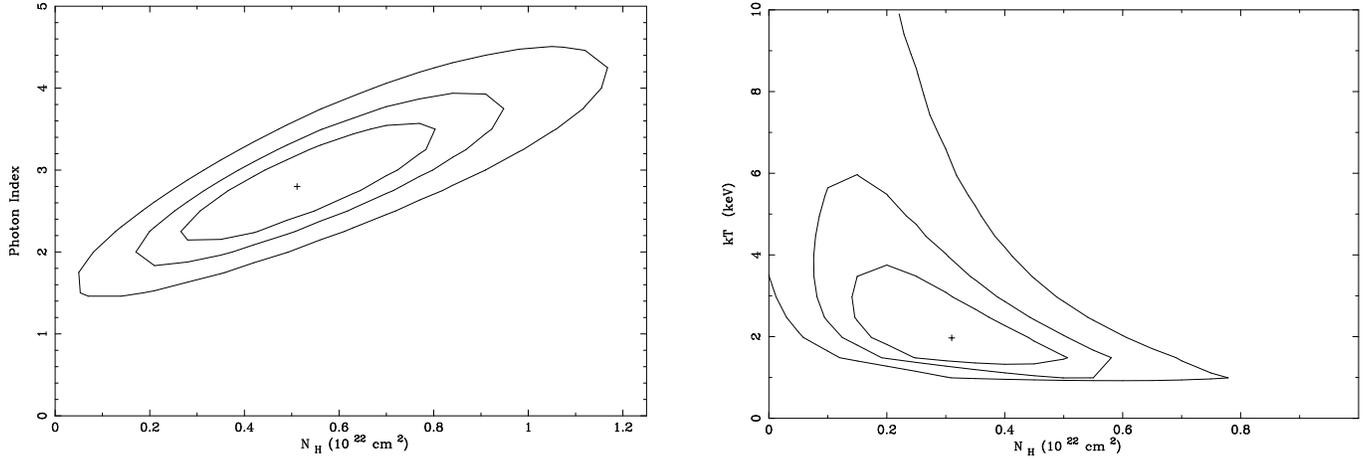

\centerline{
\includegraphics[height=8.5cm,angle=-90]{f2a.eps}\hfill
\hglue0.2in{\includegraphics[height=8.5cm,angle=-90]{f2b.eps}}}
\figcaption{({\it Left}) $N_{\rm H}$ vs. $\Gamma$ and ({\it Right})
$N_{\rm H}$ vs. $kT$ confidence contours for SBS~0335$-$052 West
\#1. Contours delineate the 66\%, 95\%, and 99\% confidence regions.
\label{fig:conf}}
\end{figure*}

The spectrum of the brighter component point source (\#1) in
SBS~0335$-$052W is acceptably fit by both the absorbed power-law or
thermal plasma models. The best fit parameters are
$N_{\rm HI}=(5.2^{+3.3}_{-2.7})\times10^{21}$~cm$^{-2}$, 
$\Gamma=2.8^{+0.9}_{-0.8}$ and
$N_{\rm HI}=(3.1^{+2.3}_{-1.9})\times10^{21}$~cm$^{-2}$, 
$kT=2.0^{+2.2}_{-0.8}$ keV, 
respectively (see Table~\ref{tab:sources}). The $N_{\rm H}$ vs.
$\Gamma$ and $N_{\rm H}$ vs. $kT$ confidence contours for 
SBS~0335$-$052W \#1 are given in
Figure~\ref{fig:conf}, indicating the relative accuracy of the derived
parameters.  The soft intrinsic spectrum is similar to that of the
point source in SBS~0335$-$052, and is not consistent with the typical
hot gas component seen in other galaxies.  The best fit values of
$N_{\rm H}$ in both models are higher than the beam-diluted $N_{\rm
  HI}= 1.0\times10^{21}$~cm$^{-2}$ obtained from the VLA HI map. However,
they are lower than the value of 9.5$\times10^{21}$~cm$^{-2}$ obtained
from scaling the $N_{\rm HI}$ obtained from the fit of the damped
L$\alpha$ profile in SBS~0335$-$052 by the ratio of radio HI column
densities of 1.35 between SBS~0335$-$052 and SBS~0335$-$052W
\citep{Pustilnik2001}.  The point source in SBS~0335$-$052W has an
unabsorbed X-ray luminosity in the range
(5.2--8.5)$\times10^{39}$~erg~s$^{-1}$.

The spectrum of the fainter component point source (\#2) in
SBS~0335$-$052W is also acceptably fit by both the absorbed power-law
or thermal plasma models. The best fit parameters are
$N_{\rm H}=2.3 (<7.1)\times10^{21}$~cm$^{2}$, 
$\Gamma=1.9^{+1.1}_{-0.8}$ and
$N_{\rm H}=1.3 (<3.0)\times10^{21}$~cm$^{2}$, 
$kT=5.4 (>1.9)$ keV, 
respectively (see Table~\ref{tab:sources}).  Although the column
density is not well constrained, the upper limits are in good
agreement with those from point source \#1.  The intrinsic spectrum of
this source is the hardest of the three discussed so far, but it is still
relatively soft.  This point source has an unabsorbed X-ray luminosity
in the range (2.4--2.8)$\times10^{39}$~erg~s$^{-1}$. 

The spectrum of I~Zw~18 has much better photon statistics than
SBS~0335$-$052, so we grouped the X-ray spectra in 20-count bins and
fit the data using $\chi^2$. The absorbed power-law model provides a
slightly better fit to the continuum than the absorbed thermal plasma
model, although both are formally acceptable. The best fit parameters are
$N_{\rm H}=(1.44^{+0.38}_{-0.37})\times10^{21}$~cm$^{-2}$, 
$\Gamma=2.01^{+0.14}_{-0.16}$ and
$N_{\rm H}=(0.87^{+0.27}_{-0.24})\times10^{21}$~cm$^{-2}$, 
$kT=4.06^{+1.84}_{-1.19}$ keV, 
respectively (see Table~\ref{tab:sources}). Interestingly, we find
large residuals below $\approx$~1~keV, particularly around
$\approx$~0.65~keV, which is most likely caused by one or more
emission lines.  \citet{Bomans2002} also found this line, accounting
for about 3\% of the total flux, which they attributed to the O{\sc
VIII} hydrogen-like emission line. 


\begin{figure*}
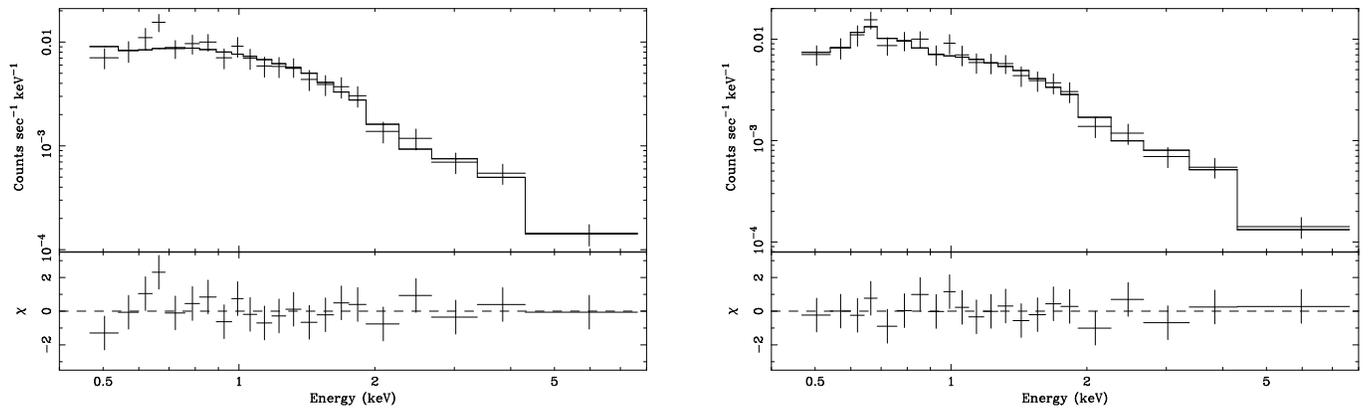

\centerline{
\includegraphics[height=8.5cm,angle=-90]{f3a.eps}\hfill
\includegraphics[height=8.5cm,angle=-90]{f3b.eps}}
\figcaption{
({\it Left}) X-ray spectrum of I~Zw~18 point source, modeled with an
absorbed power-law of slope $\Gamma=2.01$ and $N_{\rm 
H}=1.44\times10^{21}$ cm$^{-2}$. Note the large residuals around 0.65~keV
suggesting the presence of an O{\sc VIII} hydrogen-like emission
line. ({\it Right}) Same spectrum,
modeled with an absorbed thermal plasma of $kT=4.28$ keV, 
$N_{\rm H}=9.4\times10^{20}$ cm$^{-2}$, and an O abundance of $Z^{\rm
  O}=7.0Z^{\rm O}_{\odot}$, which significantly improves the fit in the  
0.65~keV region.\label{fig:spectra}}
\vspace{-0.3cm}
\end{figure*} 

One way to account for this particular emission feature is an O enhancement in the X-ray gas. If we model the spectrum with an absorbed thermal plasma 
model, varying the O abundance but keeping all other elements fixed to 
$Z=0.02Z_{\odot}$, we find an improvement over the simple thermal model 
fit at the 99.99\% confidence level (Figure~\ref{fig:spectra}). 
The best fit parameters are   

%
$N_{\rm H}=(0.94^{+0.35}_{-0.24})\times10^{21}$~cm$^{-2}$, 
$kT=4.28^{+2.25}_{-1.31}$ keV, 
$Z^{\rm O}=7.0^{+12.2}_{-4.3}Z^{\rm O}_{\odot}$, 
(see Table~\ref{tab:sources}). 


To check for the possibility that $\alpha$-elements other than O may also 
be contributing to the shape of the X-ray spectrum, we have also investigated 
a model where the abundances of the $\alpha$-elements Ne, Si, S and Ar 
vary in concert with the O abundance, but in such a way that the 
$\alpha$-element-to-oxygen abundance ratios Ne/O, Si/O, S/O and Ar/O remain 
constant, equal to the values derived by \citet{Izotov1999b} for 
low-metallicity BCDs (see their Table 6). This model gives a worse fit to the 
X-ray spectrum. We conclude that the O abundance of the X-ray gas 
in the point source in
I~Zw~18 is likely to be enhanced by a factor of $\sim$ 350 with respect
to the O abundance determined for the HII region gas, 
while the other $\alpha$-elements have abundances that are consistent with no 
enrichment with respect to the abundances in the HII region.    
 

The point source in I~Zw~18 has an unabsorbed X-ray luminosity between
1.3--1.5$\times10^{39}$~erg~s$^{-1}$, a factor of 2--4 times smaller
than those found for SBS~0335$-$052.  In addition to the point sources
discussed previously, both SBS 0335--052 and I Zw 18 also show faint
extended sources.  The flux of the extended source in SBS 0335--052 is
only about 10\% of the 0.5--10 keV flux of the point source, and that
of the one in I Zw 18 is only about 2.7\% of the flux of the point
source (Table 1).  The physical nature of these extended sources will
be discussed more in detail in the next section.

\section{Results and Discussion}\label{discussion}

Our main findings are the following: 

1) X-ray emission is detected from both SBS~0335$-$052 and
SBS~0335$-$052W, and from I~Zw~18. The 0.5--10.0~keV luminosities of
the three BCDs are in the range 1.3--8.5 $\times$ 10$^{39}$ erg s$^{-1}$. 
They are dominated by point sources, although at the
distances of these objects and with the limited statistics, it is
difficult to rule out emission from a collection of fainter point or
compact extended sources. If these sources are single objects, their 
luminosities would place them in the range of the so-called ultraluminous
X-ray sources \citep[ULXs; e.g., ][]{Makishima2000}

The X-ray spectra of the point sources are well fitted by moderately
soft power-laws, typical of X-ray emission from HMXBs or ULXs. The
high X-ray luminosities of these sources could be the result of the
lower metallicities of the BCDs studied here (if accretion is via
stellar winds), or beaming \citep[as is thought to occur in some ULXs;
e.g.,][]{King2001}. The mean luminosity of galactic HMXBs is $\sim$
5$\times$10$^{36}$~erg~s$^{-1}$ \citep{vanParadjis1995}. The X-ray
sources in SBS~0335$-$052, SBS~0335$-$052W and I~Zw~18 are
respectively about 600, 1300 and 300 times brighter. For comparison,
the HMXBs in the Magellanic Clouds (the LMC and the SMC have
respectively 1/3 and 1/8 of the Sun's metallicity) are about 50 times
brighter than their galactic counterparts \citep{vanParadjis1995}.

In the event of X-ray emission produced by accretion onto a compact
object by a stellar wind, there are two main reasons which may explain
the boosting of the X-ray luminosities of HMXBs in low-metallicity
systems \citep{vanParadjis1995}.  First, the X-ray heating of the gas
as it falls toward a compact object depends strongly on the atomic
number Z, since the photoelectric cross-section goes as Z$^4$. Thus in
low-metallicity systems, the X-ray heating is less, resulting in a
larger accretion rate because it is less impeded by heating. Second,
the accretion rate dM/dt increases strongly as the stellar wind
velocity v of an OB star as the orbit of the compact object decreases
(dM/dt is proportional to v$^{-4}$). It is known that the terminal
wind velocity in low-metallicity systems drops significantly.
\citet{Thuan1997b} have measured a terminal velocity of $\sim$ 500 km
s$^{-1}$ for SBS~0335$-$052, as compared to 2000--4000 km s$^{-1}$ for
the Galaxy and the LMC, and 1200--1500 km s$^{-1}$ for the SMC.  Thus
the high X-ray luminosities of the sources in the metal-deficient
galaxies SBS~0335$-$052, SBS 0335--052W and I~Zw~18 (i.e., $L_{\rm
  X}>10^{39}$~erg~s$^{-1}$) do not necessarily require extreme black
hole masses.

The X-ray luminosities found here are higher than the point-source
luminosities observed in less metal-poor dwarf starburst dwarf
galaxies such as NGC~3077 \citep[Z$_\odot$, $L_{\rm
  0.3--8.0~keV}\approx2$--$5\times10^{38}$, $\ga15$\% from point
sources;][]{Ott2003}, IC~10 \citep[Z$_\odot$/4, $L_{\rm
  0.3--8.0~keV}\approx2\times10^{38}$, $\ga95$\% from a single point
source;][]{Bauer2003}, NGC~4449 \citep[Z$_\odot$/4, $L_{\rm
  0.3--8.0~keV}\approx2\times10^{39}$, $\sim60$\% from point
sources;][]{Summers2003} and NGC~1569 \citep[Z$_\odot$/5, $L_{\rm
  0.3--6.0~keV}\approx1\times10^{39}$, $\sim30$\% from point
sources;][]{Martin2002, Heike2003}.  They are comparable to the point
source luminosity in Holmberg~II \citep[Z$_\odot$/10, $L_{\rm
  0.1--10.0~keV}\approx1\times10^{40}$, $\ga75$\% from a single point
source;][]{Miyaji2001}.

Interestingly, 
\citet{Hunt2003b} have detected a non-thermal radio source associated 
with SSCs 1 and 2. Its flux density is $\sim$ 0.6 mJy
at 5 GHz, corresponding to a non-thermal radio luminosity of 
$\sim$ 2 $\times$ 10$^{20}$ W Hz$^{-1}$. Within the uncertainties in
the X-ray and radio positions, the non-thermal radio source can be 
associated with the X-ray point source. However the nature of such a 
radio+X-ray emitting object is not clear. While its X-ray luminosity 
is similar to that expected from a microquasar associated with a medium-mass 
black hole, its radio luminosity is about a million times greater than that 
observed in such objects \citep{Mirabel1999}. It can also be that the X-ray 
and radio sources are physically distinct objects, although they are both 
associated with the massive starburst episode that gave birth to the SSCs.


We note also that the time scale for HMXB formation of 3--10~Myr after
the onset of the starburst is consistent with the ages of 5~Myr or
younger of SSCs 1 and 2 in SBS~0335$-$052 \citep{Vanzi2000} and of
$\sim$ 3~Myr of the NW component in I~Zw~18 \citep{Hunt2003a}.

2) In addition to the point source emission in SBS 0335--052 and I Zw 18, 
there appears also to be
faint extended diffuse X-ray emission from hot gas, hints of which can
be seen in the smoothed X-ray contours of Figure~\ref{fig:overlay},
particularly in I~Zw~18. This emission, if truly diffuse, can be
understood in both BCDs as produced by the hot gas filling the cavity
of a superbubble carved out in the ISM of the BCD by stellar winds and
supernovae.  Indeed, {\it HST} optical imaging by TIL shows a
large superbubble cavity of radius $\sim$ 380 pc in SBS~0335$-$052, 
northeast of SSC 3 (see their Figure 1),
where the diffuse X-ray emission appears to lie.  Using equations 2, 4
and 6 of \citet{Martin1995}, the X-ray luminosity $L_{\rm X}$
of the superbubble is proportional to $R^{1.65}n^{0.66}T^{2.66}$,
where $R$ is the radius of the superbubble, $n$ the ambient ISM
density and $T$ the gas temperature. Adopting $R = 380$~pc, $n =
12$~cm$^{-3}$ and $T = 7.6\times10^{6}$~K, the predicted $L_{\rm X}$
is in good agreement with the observed diffuse X-ray luminosity of
$6.4\times10^{38}$~ergs~s$^{-1}$. As for I~Zw~18, the diffuse X-ray
emission lie between the NW and SE centers of star
formation. {\it HST} optical imaging by \citet{Hunter1995}
 suggests the presence of a supernova cavity there of radius 120 pc 
(see their figure 3).  
With $R = 120$~pc, $n = 12$~cm$^{-3}$ and $T = 6.0\times10^{6}$~K, 
$L_{\rm X}$ is
equal to $5.3\times10^{37}$~ergs~s$^{-1}$, again in good agreement
with the observed diffuse X-ray luminosity in I~Zw~18.

3) The spectrum of I~Zw~18 shows a strong emission line at $\sim$ 0.65
KeV, identified by \citet{Bomans2002} to be a O VIII hydrogen-like
line. The best spectral fit (Figure~\ref{fig:spectra}) gives an
overabundance of oxygen by a factor of $\sim$ 7 with respect to the
Sun and of $\sim$ 350 with respect to the other heavy elements. It is
interesting that this line is so strong because {\it FUSE}
observations of the HI interstellar gas in I~Zw~18 \citep{Aloisi2003,
  Lecavelier2003} shows the ISM to be as or more metal-deficient than
the HII gas, i.e. to be less than 2\% of the solar metallicity. The
presence of a sizable amount of oxygen in the X-ray spectrum of the
point source is most likely due to oxygen-enrichment by either a
companion star (in the case of an X-ray binary) or undiluted supernova ejecta
(in the case of hot gas).

Further progress in our understanding of the X-ray emission in
SBS~0335$-$052, SBS~0335$-$052W and I~Zw~18 will require substantially
better photon statistics and higher spatial resolution. A long
observation with {\it XMM-Newton}, or {\it Constellation-X}/{\it XEUS}
in the future, would be able to provide better spectral and temporal
constraints that might confirm the nature of the brightest point
sources as true single compact objects or a collection of X-ray
binaries and hot gas. However, such observations will not have the
necessary spatial resolution for separating the different emission
components seen by {\it Chandra}. Spatially resolving these objects
with good sensitivity is beyond the capability of {\it Chandra} and
must await the launch of {\it Generation-X} or a similar
high-resolution X-ray observatory.

Putting our results in a cosmological context, we expect that the X-ray 
emission from high-reshift primordial galaxies to come mainly from 
very high luminosity HMXBs. 

\acknowledgements

TXT acknowledges the partial financial support of CXC grant
GO0-1149X. FEB thanks the hospitality of the Astronomy Department of
the University of Virginia and acknowledges the financial
support of CXC grant GO3-4112X and STScI grant HST-GO-09683.01A.

\end{document}